\documentclass[]{aa}
\usepackage{graphicx}
\usepackage{adjustbox}
\usepackage{txfonts}
\usepackage{url}
\makeatletter
\renewcommand*\aa@pageof{, page \thepage{} of \pageref*{LastPage}}
\makeatother
\usepackage{hyperref}
\hypersetup{
    colorlinks=true,
    linkcolor=blue,
    filecolor=magenta,      
    urlcolor=blue,
    citecolor=blue,
    }
\begin{document} 

\title{The Large Magellanic Cloud through the lens of the \textbf{{\textit{James Webb}}} Space Telescope: Binaries and the mass function in the galaxy's outskirts}
\subtitle{}

\author{M.\,V.\,Legnardi\inst{1} \and F.\,Muratore\inst{1} \and A.\,P.\,Milone\inst{1,2} \and G.\,Cordoni\inst{3} \and E.\,Dondoglio\inst{4} \and L.\,N.\,Gorza\inst{1} \and A.\,Bellini\inst{5} \and F.\,Calura\inst{6} \and S.\,Jang\inst{7} \and H.\,Jerjen\inst{3} \and A.\,Karakas\inst{8} \and E.\,P.\,Lagioia\inst{9} \and C.\,Li\inst{10,11} \and A.\,Mastrobuono-Battisti\inst{1,2,12} \and M.\,Tailo\inst{13} \and E.\,Vesperini\inst{14} \and E.\,Bortolan\inst{1} \and A.\,F.\,Marino\inst{2} \and S.\,Di\,Stefano\inst{15,16} }

\institute{Dipartimento di Fisica e Astronomia ``Galileo Galilei'', Univ. di Padova, Vicolo dell'Osservatorio 3, Padova, IT-35122 \\ \email{mariavittoria.legnardi@unipd.it}
\and
INAF - Osservatorio Astronomico di Padova, Vicolo dell’Osservatorio 5, Padova, IT-35122
\and 
Research School of Astronomy and Astrophysics, Australian National University, Canberra, ACT 2611, Australia
\and
Physics Department, American University of Sharjah, P.O. Box 26666, Sharjah, UAE
\and 
Space Telescope Science Institute, 3700 San Martin Drive, Baltimore, MD 21218, USA
\and 
INAF – Osservatorio di Astrofisica e Scienza dello Spazio di Bologna, Via Gobetti 93/3, 40129 Bologna, Italy
\and
Center for Galaxy Evolution Research and Department of Astronomy, Yonsei University, Seoul 03722, Republic of Korea
\and 
School of Physics and Astronomy, Monash University, VIC 3800, Australia
\and 
South-Western Institute for Astronomy Research, Yunnan University, Kunming, 650500 P. R. China
\and 
School of Physics and Astronomy, Sun Yat-sen University, Zhuhai 519082, China
\and
CSST Science Center for the Guangdong-Hongkong-Macau Greater Bay Area, Sun Yat-sen University, Zhuhai 519082, China
\and 
Dipartimento di Tecnica e Gestione dei Sistemi Industriali, Università degli Studi di Padova, Stradella S. Nicola 3, I-36100 Vicenza, Italy
\and 
Dipartimento di Fisica e Astronomia Augusto Righi, Università degli Studi di Bologna, Via Gobetti 93/2, 40129 Bologna, Italy
\and
Department of Astronomy, Indiana University, 727 East Third Street, Bloomington, IN 47405, USA
\and 
Dipartimento di Fisica, Sezione di Astronomia, Università di Trieste, Via Tiepolo 11, I-34143 Trieste, Italy
\and
INAF- Osservatorio Astronomico di Trieste, Via Tiepolo 11, I-34143 Trieste, Italy}

\titlerunning{The binary fraction and mass function of the LMC field} 
\authorrunning{Legnardi et al.}

\date{Received 10 June 2026 / Accepted 20 July 2026}

\abstract{Nearby galaxies such as the Large Magellanic Cloud (LMC) offer an ideal laboratory to test the initial mass function under different physical conditions, but previous works have been limited by photometric depth and have therefore poorly constrained the low-mass regime. Here, we analyze ultra-deep {\it James Webb} Space Telescope observations of a field in the LMC outskirts, near the intermediate-age and massive star cluster NGC\,1846. Using the $m_{\rm F322W2}$ versus\,$m_{\rm F115W}-m_{\rm F322W2}$ color-magnitude diagram, we derive the mass function (MF) down to unprecedentedly low masses ($M=0.17\,M_{\odot}$), explicitly accounting for the contribution of unresolved binaries, whose fraction is constrained directly from the data. For systems with mass ratios $q>0.6$, we measure a binary fraction of $f_{\rm bin}^{q>0.6}=0.15\pm0.01$, implying a total binary fraction of $f_{\rm bin}^{\rm TOT}=0.34\pm0.02$ for a flat mass-ratio distribution. This is consistent with values in the Small Magellanic Cloud (SMC) and in the Milky Way field, suggesting similar binary formation efficiency across low-density environments. We also derive the MF over the mass interval 0.17-0.82\,$M_{\odot}$ and fit it with a power law, obtaining a slope of $\alpha = -1.49 \pm 0.16$. This slope is shallower than the canonical Salpeter value ($\alpha=-2.35$) and slightly shallower than that measured in the SMC field, while remaining consistent with determinations for Galactic open clusters and for several clusters in the Magellanic Clouds and the Milky Way. Together, these results support a scenario in which both binary formation efficiency and the shape of the low-mass MF depend only weakly on the environment.}

\maketitle

\section{Introduction}
\label{sec:intro}
The stellar initial mass function (IMF) quantifies the distribution of stellar masses at birth and is a cornerstone of stellar population studies. By setting the relative numbers of low- and high-mass stars, it shapes the chemical enrichment of the interstellar medium, the supernova rate, and the integrated light of galaxies across cosmic time. Since the pioneering work of \citet{salpeter1955} and later parameterizations by \citet{kroupa2001} and \citet{chabrier2003}, the IMF has often been assumed to be approximately universal, at least within the Milky Way and nearby star-forming regions \citep[see][for reviews]{bastian2010, offner2014}.

The extent to which this assumption holds remains actively debated. On galaxy-wide scales, studies combining simulations and observations have suggested top-heavy (shallow) IMFs in dense star-forming regions \citep[e.g.,][]{abel2002,bromm2002,pouteau2022} and in high-redshift galaxies \citep[e.g.,][]{calura2009,calura2014}. Conversely, analyses based on integrated light and stellar kinematics have provided evidence for bottom-heavy (steep) IMFs in massive early-type galaxies \citep[e.g.,][]{vandokkum2010,conroy2012,cappellari2012}. On smaller scales, the most direct constraints come from star counts in resolved stellar systems, such as Galactic globular clusters \citep[GCs; e.g.,][]{paust2010,sollima2017,dondoglio2022,baumgardt2023,marino2024a} and young Milky Way open clusters \citep[e.g.,][]{cordoni2023,marchuk2026}. However, these studies are predominantly limited to the Milky Way, restricting the range of environmental conditions over which the IMF can be robustly tested.

Resolved stellar populations in nearby dwarf galaxies offer a key opportunity to test the IMF over a wider range of metallicities, star formation histories, and structural properties. The shallow galactic gravitational potential and their long dynamical timescales imply that present-day mass functions (MFs) are only mildly affected by dynamical evolution, retaining a closer imprint of the IMF \citep[][and references therein]{geha2013}. Observations with the \textit{Hubble} Space Telescope (HST) in ultra-faint dwarfs (UFDs) and the Magellanic Clouds have indicated MFs shallower than the Salpeter slope ($\alpha=-2.35$). For instance, \citet{wyse2002} measured $\alpha \sim -1.8$ in the Ursa Minor dwarf galaxy, consistent with results for other UFDs \citep[][]{geha2013,gennaro2018a,gennaro2018b}, while \citet{kalirai2013a} reported $\alpha \sim -1.9$ in the Small Magellanic Cloud (SMC) field. However, these constraints typically do not reach below $\sim 0.4\,M_{\odot}$, leaving the low-mass regime of the IMF largely unexplored.

The {\it James Webb} Space Telescope (JWST) has opened access to this parameter space. Its unprecedented sensitivity in the near-infrared, combined with its high angular resolution, enables star counts well below $0.4\,M_{\odot}$ in nearby galaxies, allowing direct measurements of the low-mass IMF in extragalactic environments for the first time. In a recent study, we exploited these capabilities to investigate an SMC field ($\alpha_{\rm J2000} \sim 00^{\rm h}21^{\rm m}16^{\rm s}$, $\delta_{\rm J2000} \sim -72^{\rm d}06^{\rm m}16^{\rm s}$) near the GC 47\,Tucanae, at a projected distance of $\sim 2.5^{\circ}$ (2.7\,kpc) from the SMC center \citep[][]{legnardi2025}. Using the $m_{\rm F322W2}$ versus $m_{\rm F115W}-m_{\rm F322W2}$ color–magnitude diagram (CMD), we simultaneously constrained the fraction of unresolved binary systems and derived the MF down to $\sim 0.2\,M_{\odot}$, obtaining $\alpha = -1.99 \pm 0.08$. More recently, \cite{cohen2026} analyzed JWST observations of a different SMC field ($\alpha_{\rm J2000} \sim 01^{\rm h}00^{\rm m}36^{\rm s}$, $\delta_{\rm J2000} \sim -74^{\rm d}59^{\rm m}42^{\rm s}$), measuring the MF down to $\sim 0.16\,M_{\odot}$ and finding $\alpha=-1.61\pm0.03$. These results highlight both the diagnostic power of JWST and the remaining uncertainties in the determination of the IMF slope at low masses in dwarf galaxy environments.

In this work, we extend the same methodology to the Large Magellanic Cloud (LMC) using deep observations obtained with the Near-Infrared Camera (NIRCam) on board the JWST. We focus on a field located in the outskirts of the LMC, in the vicinity of the intermediate-age cluster NGC\,1846. The LMC is an especially valuable laboratory, thanks to its rich cluster system and extended field population tracing a complex formation history. Previous IMF constraints in LMC clusters, star-forming regions, and field populations \citep[e.g.,][]{holtzman1997,gouliermis2005,gouliermis2006,dario2009,liu2009a,liu2009b,kalari2018} were limited by photometric depth and spatial resolution and could not robustly probe the low-mass regime.

Here we leverage JWST depth and resolution to derive the field luminosity function and infer the stellar MF down to unprecedentedly low masses, explicitly accounting for observational effects and the presence of unresolved binaries that can significantly bias the main-sequence (MS) luminosity distribution. This provides new constraints on the IMF in an extragalactic field, and a stringent test of IMF universality.

The paper is organized as follows. In Sect.~\ref{sec:data} we describe the observations and data reduction procedures. In Sect.~\ref{sec:phot} we present the CMDs of NGC\,1846 and its surrounding field and discuss the main stellar populations. In Sect.~\ref{sec:bin} we derive the fraction of photometric binaries, and in Sect.~\ref{sec:MF} we compute the luminosity function and infer the stellar MF of the LMC field. Finally, Sect.~\ref{sec:concl} summarizes our results and conclusions.
  
%%%%%%%%%%%%%%%%%%%%%%%%%%%%%%%%%
\begin{figure*}
    \centering
    \includegraphics[width=.95\textwidth,trim={0cm 24cm 0cm 0cm},clip]{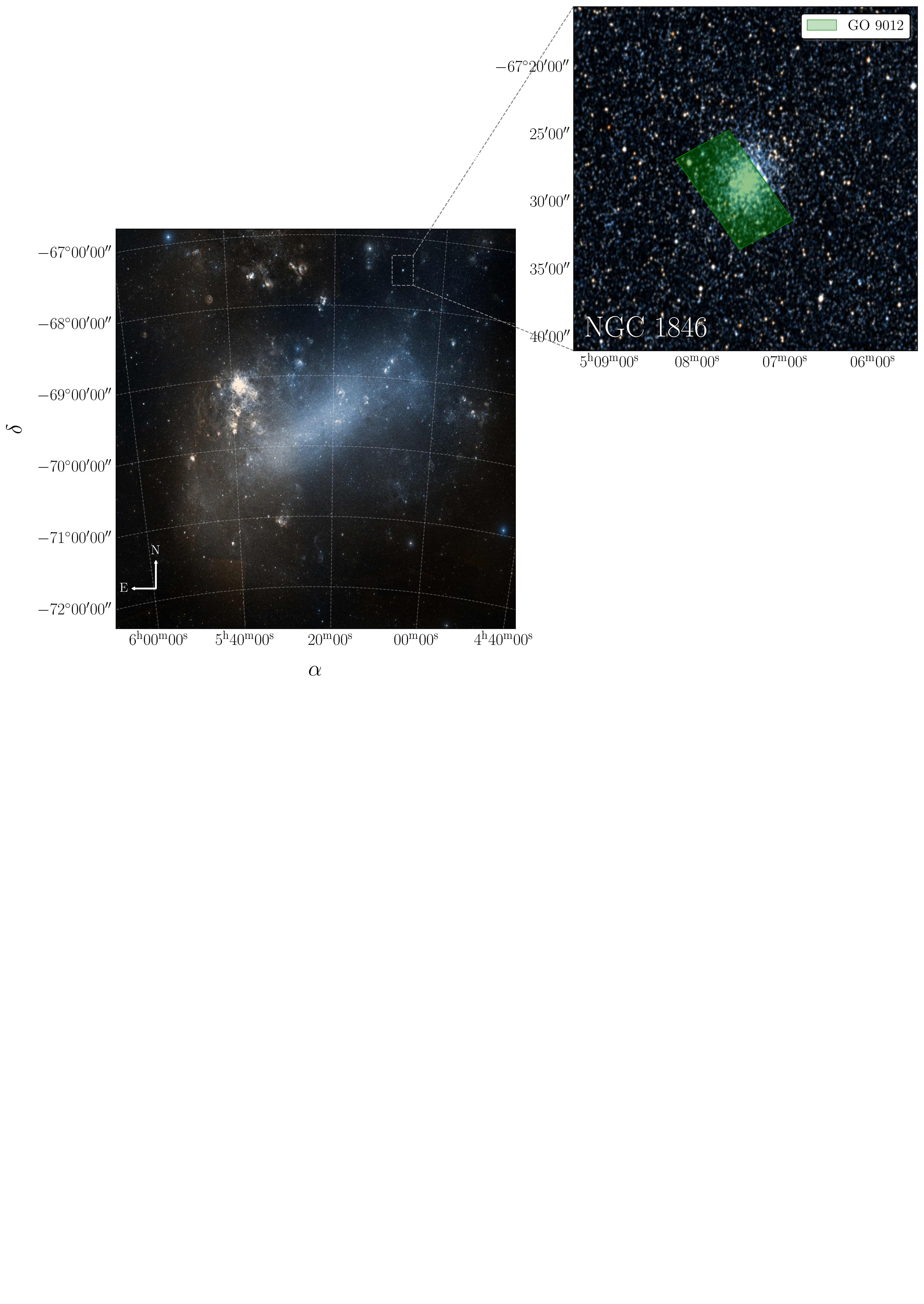}  
    \caption{Image of the LMC obtained from the Digitized Sky Survey 2. The inset shows a zoomed-in view of the region surrounding the intermediate-age cluster NGC\,1846, where the observations analyzed in this study were performed. The NIRCam field of view is outlined in green. North is up and east is to the left.}
    \label{fig:LMC_data}
\end{figure*}
%%%%%%%%%%%%%%%%%%%%%%%%%%%%%%%%%
%%%%%%%%%%%%%%%%%%%%%%%%%%%%%%%%%
\begin{figure*}
    \centering
    \includegraphics[width=.95\textwidth]{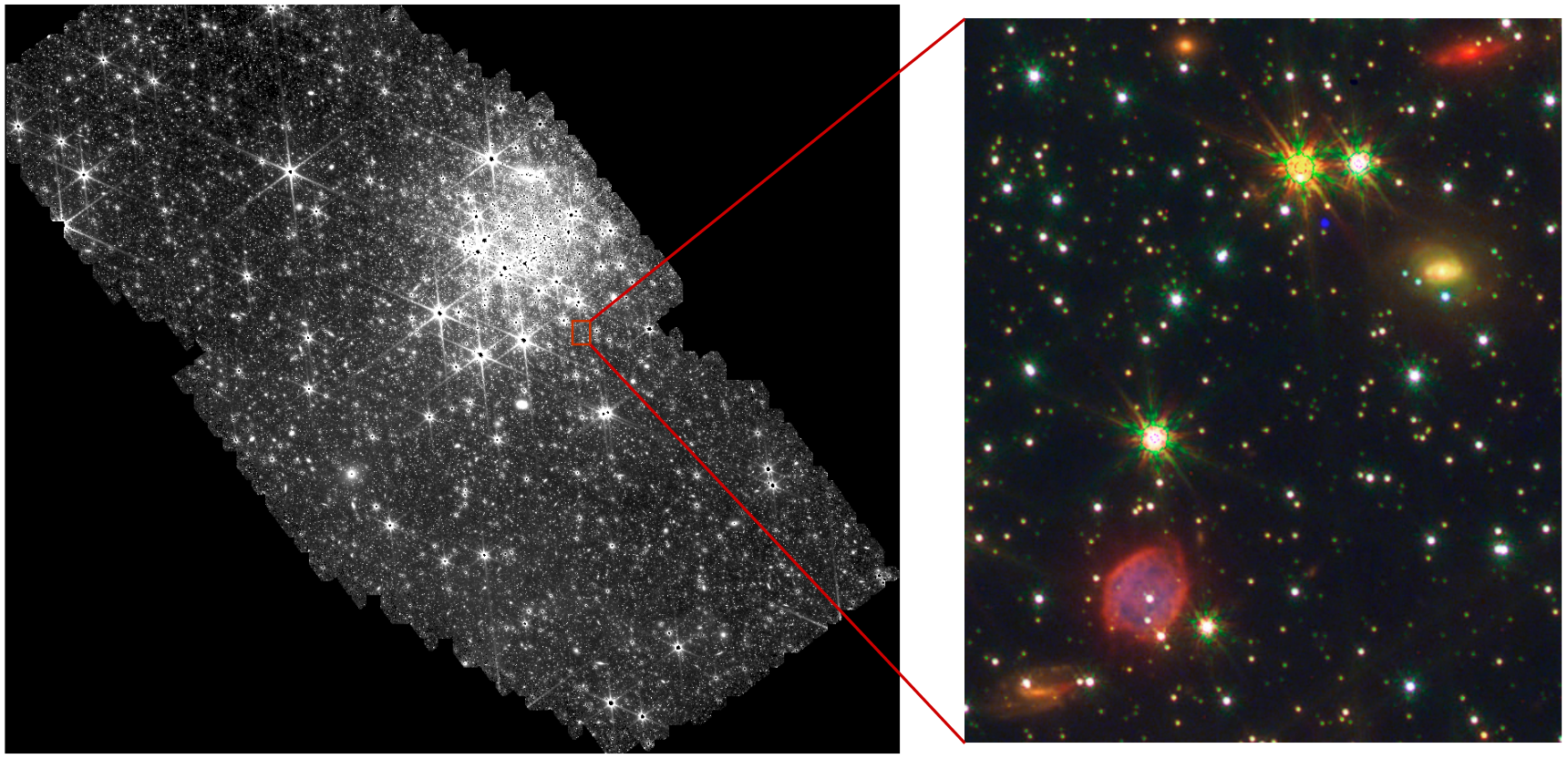}  
    \caption{Overview of the observational dataset used in this work. {\it Left panel.} Stacked NIRCam/F322W2 image of the star cluster NGC\,1846 and the surrounding LMC field. {\it Right panel.} Three-color composite zoom-in of a representative central region, with the blue, green, and red channels corresponding to the stacked F555W (HST), F115W, and F322W2 images, respectively.}
    \label{fig:NGC1846_image}
\end{figure*}
%%%%%%%%%%%%%%%%%%%%%%%%%%%%%%%%%

\section{Observations and data reduction}
\label{sec:data}
To study low-mass stars in the field of the LMC, we used deep JWST/NIRCam images obtained through the F115W and F322W2 filters. The data were collected within the GO-9012 program (PI: A.\,P.\,Milone), originally designed to investigate the presence of multiple stellar populations in the intermediate-age \citep[1.6 Gyr;][]{milone2023a} and massive \citep[$1.7\times10^5$\,$M_{\odot}$;][]{goudfrooij2014} cluster NGC\,1846 . As shown in the inset of Fig.~\ref{fig:LMC_data}, the NIRCam footprint is centered at $\alpha_{\rm J2000} \sim 05^{\rm h}07^{\rm m}43^{\rm s}$ and $\delta_{\rm J2000} \sim -67^{\rm d}28^{\rm m}28^{\rm s}$, at a projected distance of $\sim 3.0^{\circ}$ from the LMC center, which corresponds to 2.6\,kpc assuming a distance of 50.1\,kpc derived from the best-fitting isochrone.  

Figure~\ref{fig:NGC1846_image} presents the stacked F322W2 image used in our analysis, showing that the observed field covers both NGC\,1846 and the surrounding LMC field population. A three-color composite zoom-in of a representative central region is also shown, highlighting the planetary nebula Mo-17 \citep[][see also \citealt{mackey2013}]{morgan1994}, visible in the lower-left corner. The composite image was created by combining stacked F555W (from HST) data with JWST F115W and F322W2 exposures in the blue, green, and red channels, respectively. In particular, the blue-channel image is based on stacked F555W observations obtained with the Wide Field Channel of the Advanced Camera for Surveys (ACS/WFC) on board the HST and processed by \cite{milone2023a}. These data were originally collected as part of programs GO-9891 (PI: G.\,F.\,Gilmore) and GO-10595 (PI: P.\,Goudfrooij).

NIRCam consists of two channels operating simultaneously at short (SW) and long (LW) wavelengths, each comprising two modules separated by a gap of $\sim 44\arcsec$. The observations were acquired on 2025 October 20 and consist of 36 exposures per filter, each with an integration time of 1374\,s. During each exposure, the F115W and F322W2 filters were used simultaneously in the SW and LW channels, respectively. The observations employed the DEEP8 readout pattern and a ‘FULL 45’ dither pattern optimized to bridge detector gaps and improve image sampling.
 
We built the astro-photometric catalog using KS2, an advanced evolution of the \texttt{kitchen\_sync} software originally developed by \citet{anderson2008}. KS2 simultaneously fits all available exposures and provides three complementary photometric methods, tailored to different magnitude regimes \citep[see][for details]{sabbi2016,bellini2017,nardiello2018}. Since our primary goal is to measure the stellar MF of the LMC field down to very low masses, we adopted the measurements from {\it Method III}, optimized for very faint sources in moderately crowded fields. After subtracting neighbor stars, this method performs aperture photometry within a radius of 0.75 pixels and estimates the local sky from an annulus with inner and outer radii of 2 and 4 pixels, respectively.

KS2 outputs a set of diagnostics to quantify measurement quality. We selected well-measured, isolated sources following the criteria described by \citet[][see their Sect.~2.4]{milone2023a}. Instrumental magnitudes were calibrated to the Vega system as in \citet{milone2023a}, including encircled-energy corrections and NIRCam photometric zero-points released by the Space Telescope Science Institute\footnote{\url{https://jwst-docs.stsci.edu /jwst-near-infrared-camera/nircam-performance/nircam-absolute-flux-calibration-and-zeropoints}}. We further corrected for pixel-area variations and applied geometric-distortion corrections using the NIRCam solutions provided by J.\,Anderson\footnote{\url{https://www.stsci.edu/stsci-research/research-directory/jay-anderson}}. The level of differential reddening across the field is negligible and has no measurable impact on the MF determination; consequently, no differential-reddening correction was required.

\begin{figure*}
    \centering
    \includegraphics[width=.9\textwidth]{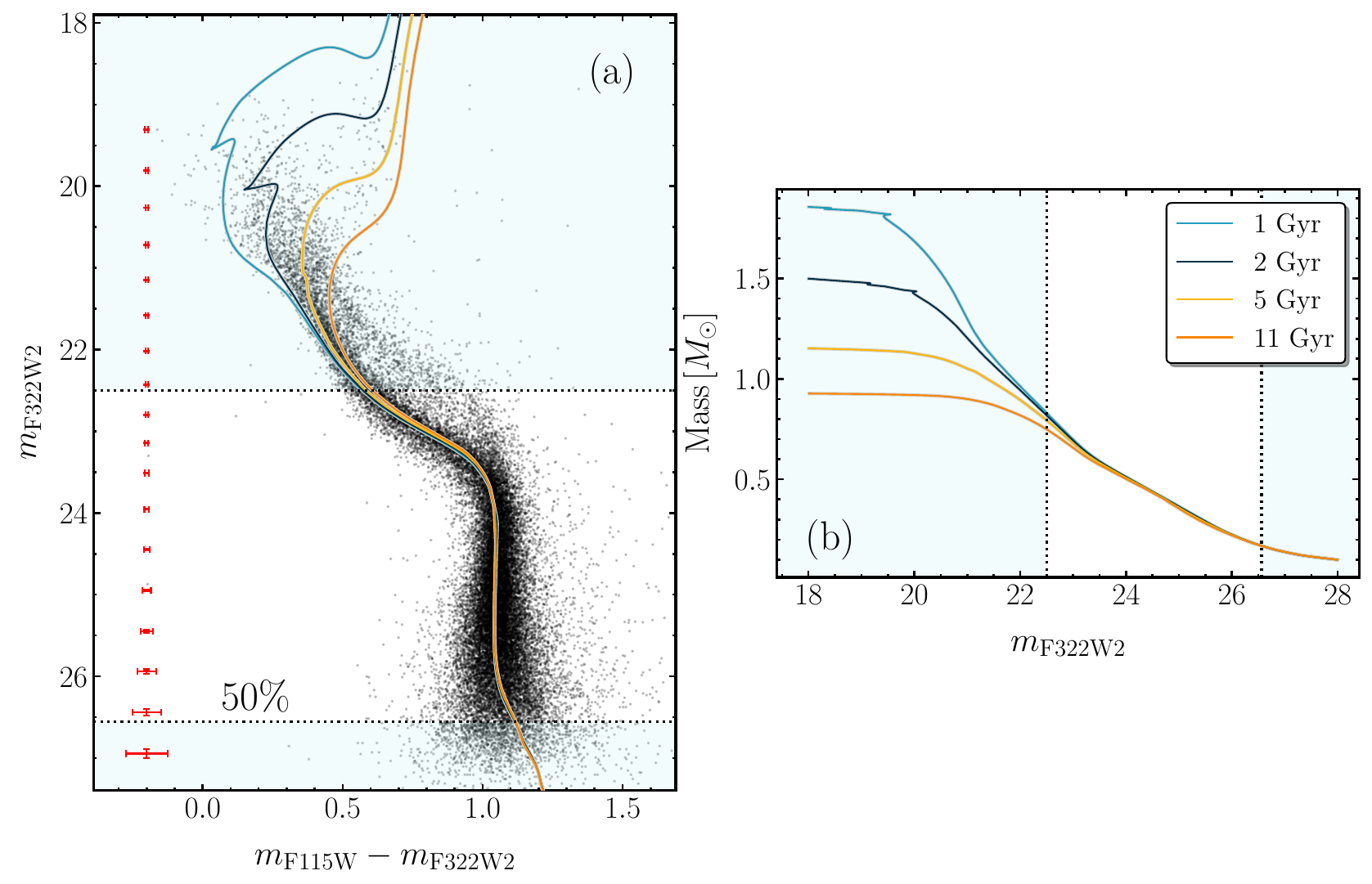}
    \caption{Illustration of the data used to investigate the LMC field. \textit{Panel a.} $m_{\rm F322W2}$ vs.\,$m_{\rm F115W}-m_{\rm F322W2}$ CMD of the LMC field. Four BaSTI isochrones with ages between 1 and 11\,Gyr are overplotted, while photometric uncertainties are indicated in the left corner. The horizontal dotted lines at $m_{\rm F322W2}=22.5$ and $m_{\rm F322W2}=26.6$ mark, respectively, the magnitude above which the mass-luminosity relation becomes strongly age-dependent and the magnitude at which the photometric completeness drops below $50\%$. \textit{Panel b.} Relation between the mass and the $m_{\rm F322W2}$ magnitude for the four isochrones shown in panel a. The vertical dotted lines indicate the same magnitude limits highlighted in panel a.} 
    \label{fig:cmd_fit}
\end{figure*}

We quantified photometric uncertainties and completeness via artificial-star (AS) tests. Following \citet{anderson2008}, we injected 10$^5$ ASs with fixed positions and fluxes, sampling the same spatial distribution as the observed stars. We assigned artificial magnitudes along the LMC MS fiducial line and processed the injected sources with KS2 using the same configuration and quality cuts adopted for real stars. ASs were added, measured, and removed one at a time by KS2, ensuring that they never interfered with one another. We computed completeness as the ratio of recovered to injected stars. The 50$\%$ completeness limit occurs at $m_{\rm F322W2} = 26.56$\,mag, as illustrated in Fig.~\ref{fig:cmd_fit}.

%%%%%%%%%%%%%%%%%%%%%%%%%%%%%%%%%
\begin{figure*}
    \centering
    \includegraphics[width=.9\textwidth]{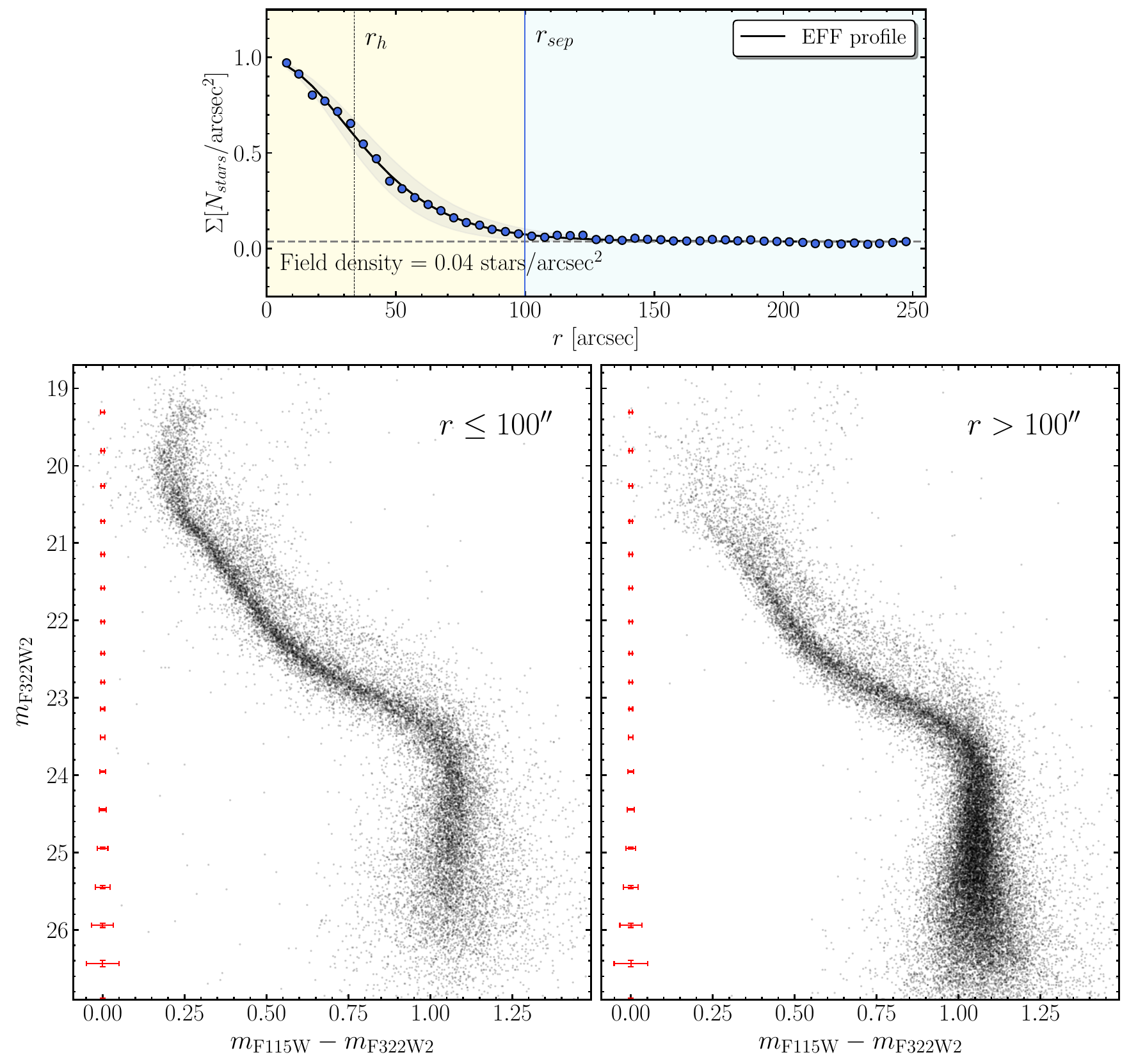}  
    \caption{$m_{\rm F322W2}$ vs.\,$m_{\rm F115W}-m_{\rm F322W2}$ CMDs for the regions at $r\le100$\,\arcsec (bottom-left panel) and $r>100$\arcsec (bottom-right panel), dominated by NGC\,1846 members and LMC field stars, respectively. The average color and magnitude uncertainties, calculated for stars in different magnitude bins, as a function of magnitude are indicated by the red error bars plotted on the left side of each diagram. The separation is based on the radial density profile of NGC\,1846, shown in the top panel. The horizontal dashed line marks the background star density, while the vertical dashed and blue solid lines indicate the half-light radius of NGC\,1846 \citep[][]{usher2017} and the adopted separation radius, respectively.} 
    \label{fig:cmd}
\end{figure*}
%%%%%%%%%%%%%%%%%%%%%%%%%%%%%%%%%
%%%%%%%%%%%%%%%%%%%%%%%%%%%%%%%%%
%%%%%%%%%%%%%%%%%%%%%%%%%%%%%%%%%

\section{The color-magnitude diagram of the Large Magellanic Cloud}
\label{sec:phot}
Figure~\ref{fig:cmd} shows the $m_{\rm F322W2}$ versus\,$m_{\rm F115W}-m_{\rm F322W2}$ CMDs for the region dominated by NGC\,1846 stars (bottom-left panel) and the surrounding LMC field (bottom-right panel). We separated the two components using the cluster radial density profile (top panel of Fig.~\ref{fig:cmd}).

To build the density profile, we divided the field of view into 50 concentric annuli of width 5\arcsec and computed the area of each annulus. We then measured the stellar surface density ($\Sigma$), as the number of stars per unit area. To reduce contamination from field stars, we considered only sources brighter than $m_{\rm F322W2} = 21.0$ mag. The resulting profile was fitted, via least squares, with an Elson-Fall-Freeman \citep[EFF;][]{elson1987} model, appropriate for young and intermediate-age Magellanic Cloud clusters that are not tidally truncated and show a power-law decline at large radii,
\begin{equation}
    \Sigma(r)=\Sigma_0\left(1+\frac{r^2}{a^2}\right)^{-\frac{\gamma}{2}}+\Sigma_{bg}
\end{equation}
where $\Sigma_0$ is the central surface density (in $N_{stars}$/arcsec$^{2}$), $a$ is the scale radius,
$\gamma$ sets the outer slope, and $\Sigma_{bg}$ is the residual background level. 

The best-fit EFF profile declines to the background level ($\Sigma_{bg}=0.04$\,stars/arcsec$^2$) at a radius of $\sim 100\arcsec$, which was therefore adopted as the boundary separating the cluster-dominated region from the region dominated by LMC field stars. This value corresponds to approximately three times the half-light radius of NGC\,1846 \citep[$r_h=34\arcsec$;][]{usher2017}. By integrating the cluster component of the best-fitting EFF profile beyond 100\arcsec, we estimated that the residual contribution from NGC\,1846 stars in the selected LMC field region is $\sim 2\%$ of the total stellar population. This confirms that the adopted radial cut effectively minimizes contamination from cluster members.

To verify that the binary fraction and MF slope of the LMC field are not affected by this choice, we repeated the analysis using different radial cuts in the range 90-$125\arcsec$, in steps of $5\arcsec$, sampling the transition region between cluster-dominated and field-dominated regimes. We verified that varying the separation radius does not significantly affect either the binary fraction or the MF slope derived for the LMC field. 

Both CMDs show a well-defined MS, extending from the turn-off, at $m_{\rm F322W2} \sim 20.5$, down to more than two magnitudes below the MS knee ($\sim 23.5$\,mag). A distinct sequence of MS-MS binaries is visible on the red side of the MS.

In the cluster-dominated CMD we confirm the presence of an extended MS turn-off \citep[e.g.,][]{mackey2008,milone2009,kamann2020}. Since the $m_{\rm F115W}-m_{\rm F322W2}$ color is highly sensitive to oxygen variations in low-mass stars, it provides an excellent diagnostic for identifying multiple populations below the MS knee. In this region we found no significant intrinsic color spread, suggesting the absence of multiple stellar populations with distinct light-element abundances in this mass range, in agreement with results at higher masses \citep[e.g.,][]{oh2023}. A dedicated analysis of the stellar populations in NGC\,1846 will be presented in a forthcoming paper (Milone et al., in prep.).

The LMC field-dominated CMD instead reveals a composite population, especially around the MS turn-off. In this region, the observed turn-off exhibits a significant broadening in the $m_{\rm F115W}-m_{\rm F322W2}$ color that exceeds the expected spread from photometric uncertainties alone, as indicated by the error bars in the lower-left corner of Fig.~\ref{fig:cmd_fit}a. This indicates that the observed color dispersion is intrinsic to the stellar population.

In Fig.~\ref{fig:cmd_fit}a we compare the data with BaSTI isochrones \citep{pietrinferni2021} spanning ages from 1 to 11 Gyr. As reported in Table~\ref{tab1}, all models assume $Z=0.006$, $(m-M)_0=18.5$, $[\alpha/{\rm Fe}]=+0.2$, and $E(B-V)=0.01$. The turn-off region is well bracketed by the youngest and oldest isochrones, consistent with an extended star-formation history in the LMC field, which naturally produces the observed broadening.

At fainter magnitudes, isochrones of different ages progressively converge along the lower MS and become nearly indistinguishable around $m_{\rm F322W2}\sim22.5$\,mag. In this regime, the CMD position is primarily driven by stellar mass rather than age. This is illustrated in Fig.~\ref{fig:cmd_fit}b, which shows stellar mass as a function of $m_{\rm F322W2}$ for the same set of isochrones displayed in panel a.

In the following sections, we use the morphology of the upper MS to derive a photometric estimate of the binary fraction in the LMC field (Sect.~\ref{sec:bin}) and exploit the depth of the JWST data to probe the low-mass end of the field MF (Sect.~\ref{sec:MF}).

%%%%%%%%%%%%%%%%%%%%%%%%%%%%%%%%%
\begin{figure}
    \centering
    \includegraphics[width=.45\textwidth]{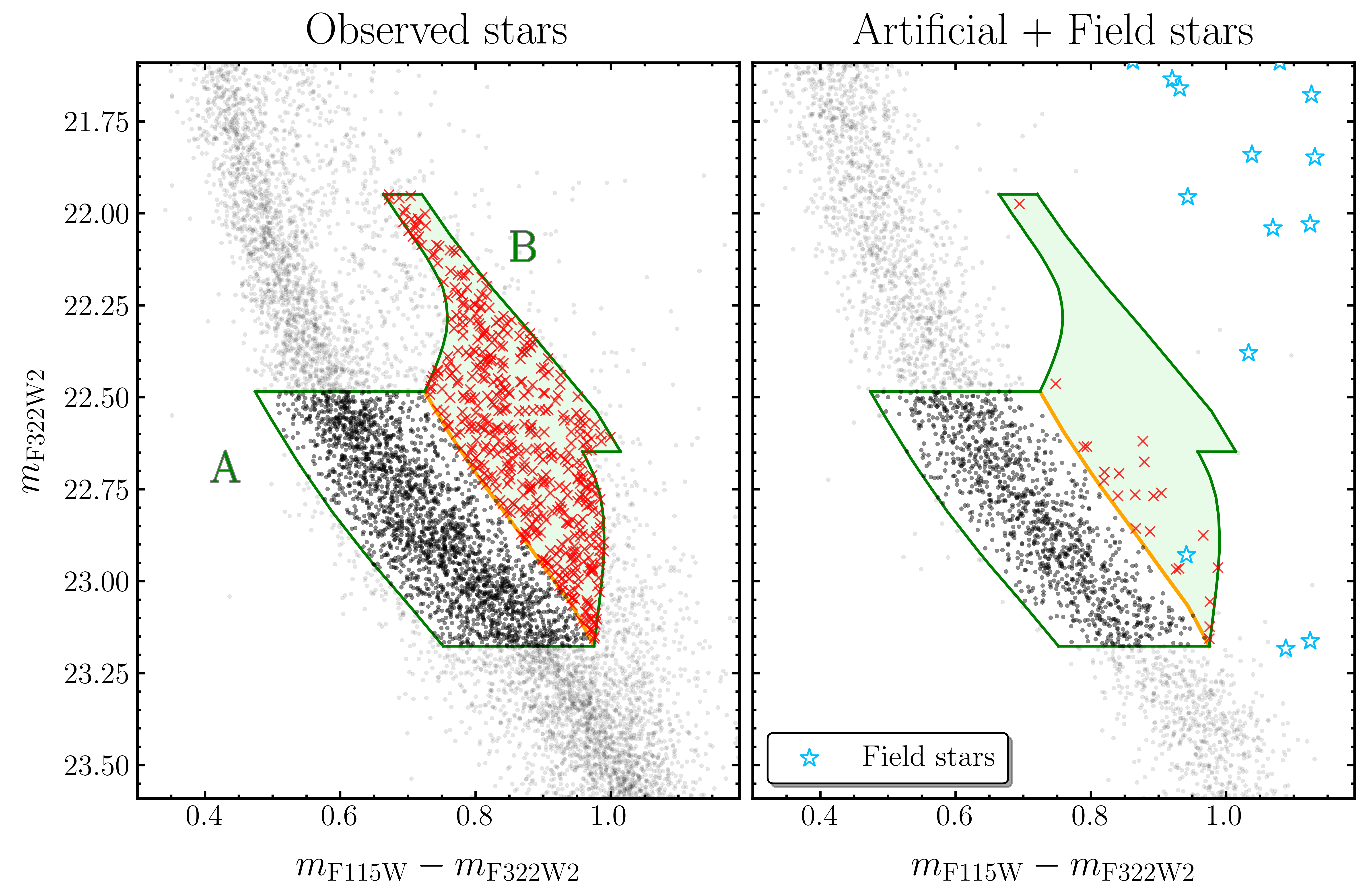} 
    \caption{Binary fraction estimation in the LMC field. {\it Left panel.} $m_{\rm F322W2}$ vs.\,$m_{\rm F115W}-m_{\rm F322W2}$ CMD of LMC stars, zoomed-in on the region used to estimate the binary fraction. {\it Right panel.} Same as the left panel but for artificial and field stars (azure star symbols). In both panels, the green lines outline region A of the CMD, adopted to derive the binary fraction. The green shaded area marks region B, a subregion of A predominantly populated by binary systems with mass ratios $q>0.6$. The orange solid line represents the fiducial sequence of binaries with $q=0.6$. Single MS stars and binary systems are shown as black points and red crosses, respectively, while all other stars are plotted in gray.} 
    \label{fig:bin}
\end{figure}
%%%%%%%%%%%%%%%%%%%%%%%%%%%%%%%%%
%%%%%%%%%%%%%%%%%%%%%%%%%%%%%%%%%
\begin{figure}
    \centering
    \includegraphics[width=.45\textwidth]{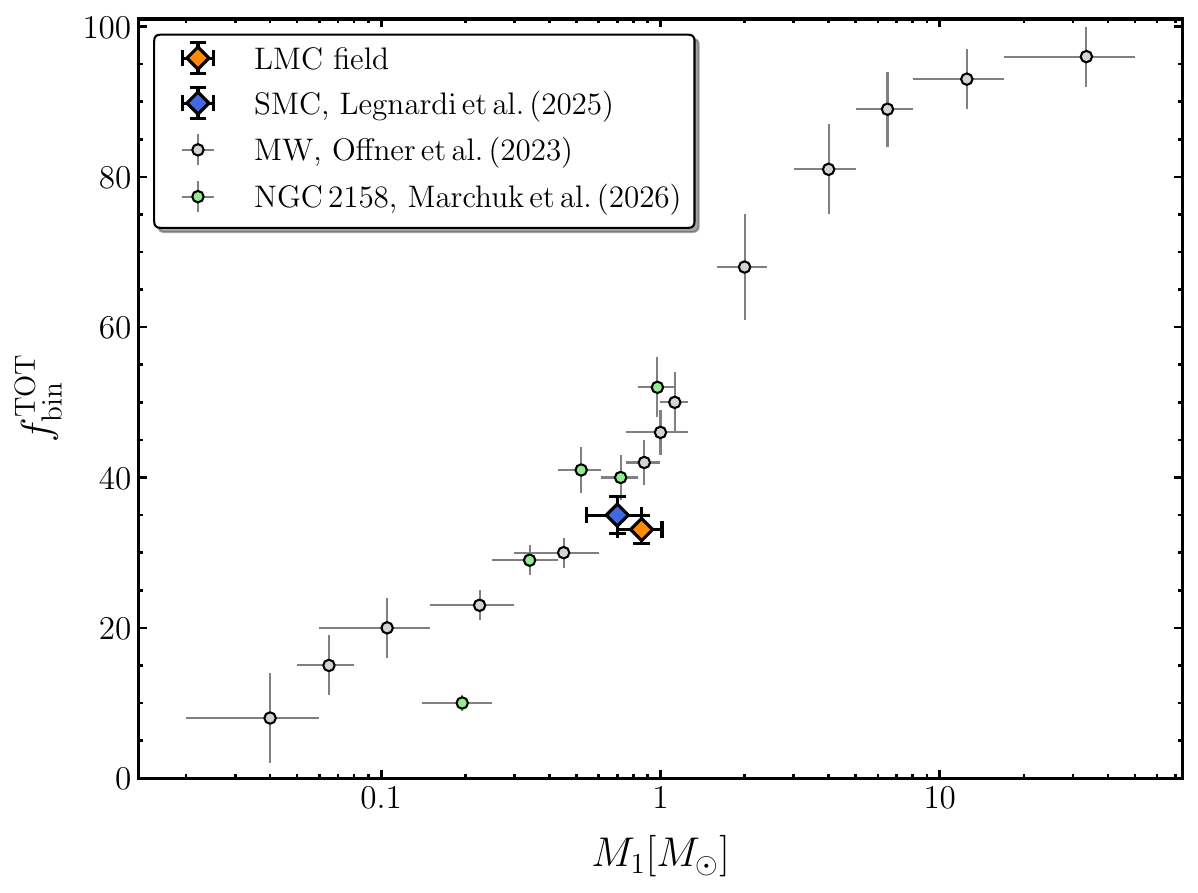}
    \caption{Total binary fraction as a function of the mass of the primary star. Gray and light green points show, respectively, Milky Way measurements compiled by \citet{offner2023} and the fractions measured for the open cluster NGC\,2158 by \citet{marchuk2026}, respectively. The orange and blue diamonds denote the total binary fractions measured for the LMC (this work) and the SMC \citep[][]{legnardi2025}, respectively.}  
    \label{fig:binplot}
\end{figure}
%%%%%%%%%%%%%%%%%%%%%%%%%%%%%%%%%

\section{Photometric binaries in the field of the Large Magellanic Cloud}
\label{sec:bin}
The depth and angular resolution of the JWST observations enable us to investigate the binary population of the LMC field over a wide range of orbital separations. While most binaries remain unresolved and can only be identified through their photometric signatures in the CMD, the widest systems can be directly resolved with NIRCam. In the following sections, we investigate both the unresolved (Sect.~\ref{subsec:unr_bin}) and resolved binary populations of the LMC field (Sect.~\ref{subsec:wide_bin}).

\subsection{Unresolved binaries in the field of the Large Magellanic Cloud}
\label{subsec:unr_bin}
Because of the large distance of the LMC ($(m-M)_0=18.5$, corresponding to $50.1$\,kpc), binary systems are unresolved and appear as single point-like sources in the $m_{\rm F322W2}$ versus\,$m_{\rm F115W}-m_{\rm F322W2}$ CMD. The magnitude of an unresolved binary system is 
\begin{equation}
    m_{\rm bin} = m_1 - 2.5 \log \left(1+\frac{F_2}{F_1}\right),
\end{equation}
where $m_1$ is the primary-star magnitude, and $F_1$ and $F_2$ are the fluxes of the primary and secondary component, respectively.

In the CMD of a simple stellar population, the position of a MS-MS binary is set by the mass of the primary star ($M_1$) and by the mass ratio, $q=M_2/M_1$. For $q\sim0$, binaries are nearly indistinguishable from single stars and lie close to the MS fiducial. For $q\sim1$, they define a sequence approximately parallel to the MS and $\sim0.75$ mag brighter, while intermediate $q$ systems populate the region in between on the red and bright side of the MS.

Since binaries with low mass ratios are indistinguishable from single stars at the LMC distance, we restricted our analysis to systems with $q>0.6$. To estimate the fraction of binaries with $q>0.6$ we followed the method introduced by \citet{milone2012a} and adopted in several previous studies \citep[e.g.,][]{milone2016a,milone2025,cordoni2023,mohandasan2024,muratore2024,muratore2026,bortolan2025,legnardi2025}.

As shown in Fig.~\ref{fig:bin}, we defined two regions well above the magnitude level below which the completeness drops below $50\%$ ($m_{\rm F322W2}=26.56$\,mag). Region A corresponds to the area enclosed by the green solid line in Fig.~\ref{fig:bin}. This region includes single MS stars with $22.5 < m_{\rm F322W2} < 23.2$ and binaries whose primary component falls in the same magnitude range. Region B (green shaded area) is the subset of A located redward of the $q=0.6$ binary fiducial (orange solid line) and is therefore expected to be dominated by binaries with $q>0.6$. Stars falling in region B are marked with red crosses, while the remaining stars in region A are shown as black points.

The fraction of binaries with $q>0.6$ is computed as
\begin{equation}\label{eq_bin}
    f_{\rm bin}^{q>0.6}=\frac{N^{B}_{\rm obs}-N^{B}_{\rm field}}{N^{A}_{\rm obs}-N^{A}_{\rm field}}-\frac{N^{B}_{\rm sim}}{N^{A}_{\rm sim}},
\end{equation}
where $N^{A,B}_{\rm obs}$ are the completeness-corrected numbers of observed sources in regions A and B, respectively. The quantities $N^{A,B}_{\rm sim}$ refer to the corresponding counts in the AS CMD, while $N^{A,B}_{\rm field}$ are the expected numbers of foreground/background contaminants. We estimated the Galactic field contamination using a simulation from the TRILEGAL code \citep{girardi2005} computed for a region with the same area and Galactic coordinates as our target. The simulated contaminants are shown as azure star symbols in the right panel of Fig.~\ref{fig:bin}.

We found a binary fraction of $f_{\rm bin}^{q>0.6}=0.15\pm0.01$ for the LMC field. Uncertainties were computed assuming Poisson statistics on the star counts entering Eq.~\ref{eq_bin} and propagated to $f_{\rm bin}^{q>0.6}$.  Assuming a flat mass-ratio distribution, i.e., a constant probability distribution for all mass ratios $0<q<1$, we inferred a total binary fraction of $f_{\rm bin}^{\rm TOT}=0.34\pm0.02$. This assumption is supported by several studies of binary populations in stellar systems \citep[][]{milone2012a,milone2016a,cordoni2023}. The derived parameters are reported in Table~\ref{tab1}.

Figure~\ref{fig:binplot} compares our total binary fraction (orange diamond) with the SMC field value from \citet[][blue diamond]{legnardi2025}, Milky Way measurements compiled by \citet[][gray points]{offner2023}, and the fractions measured in five distinct mass intervals for the open cluster NGC\,2158 \citep[][light green points]{marchuk2026}. The LMC measurement follows the same overall trend with primary-star mass, suggesting that binary formation and early evolution may be only weakly dependent on the global properties of the host galaxy in low-density stellar environments.

\subsection{Wide binaries in the field of the Large Magellanic Cloud}
\label{subsec:wide_bin}
The unprecedented sensitivity and angular resolution of the JWST and in particular of NIRCam, enable, for the first time, a direct characterization of the wide-binary population in the LMC field. This spatial resolution allows us to resolve the components of binaries at separations that were previously inaccessible, opening a new window on wide systems in external galaxies.

To estimate the fraction of resolved binaries in the LMC field, we adopted Eq.\,(2) from \cite{marchuk2026}:
\begin{equation}
    f^{q>0.6}_{\rm resolved\,bin}=\frac{N^{double}_{\rm REAL}-N^{double}_{\rm FIELD}}{N^{MS}_{\rm REAL}-N^{MS}_{\rm FIELD}}-\frac{N^{double}_{\rm AS}}{N^{MS}_{\rm AS}}.
\end{equation}
Here, $N^{double}_{\rm REAL}$ is the number of apparent double MS systems, defined as MS stars with a resolved MS companion within 5 pixels, corresponding to 7750 AU (0.16\arcsec) at the distance of the LMC, and with a companion-to-primary mass ratio $q>0.6$. The term $N^{double}_{\rm FIELD}$ accounts for contamination by chance superpositions and represents the number of doubles expected from field stars. Finally, $N^{double}_{\rm AS}$ refers to the corresponding counts derived from the AS CMD, while $N^{MS}_{\rm REAL}$, $N^{MS}_{\rm FIELD}$, and $N^{MS}_{\rm AS}$ are the numbers of observed, field, and AS MS stars with $21.5<m_{\rm F322W2}<23.0$, respectively.

We found a wide-binary fraction of $f^{q>0.6}_{\rm resolved,bin}=0.01\pm0.01$. Likewise, the fraction of systems with projected separations smaller than 2.5\,pixels (i.e., $\sim 3900$\,AU, $\sim 0.08\arcsec$) is consistent with zero within the uncertainties. These results are broadly consistent with measurements in other low-density stellar environments, including the open cluster NGC\,2158 \citep[][]{marchuk2026} and the UFDs Reticulum\,II and Bootes\,I \citep[][]{safarzadeh2022,sharjat2025}.

%%%%%%%%%%%%%%%%%%%%%%%%%%%%%%%%%
\begin{figure*}
    \centering
    \includegraphics[width=.95\textwidth]{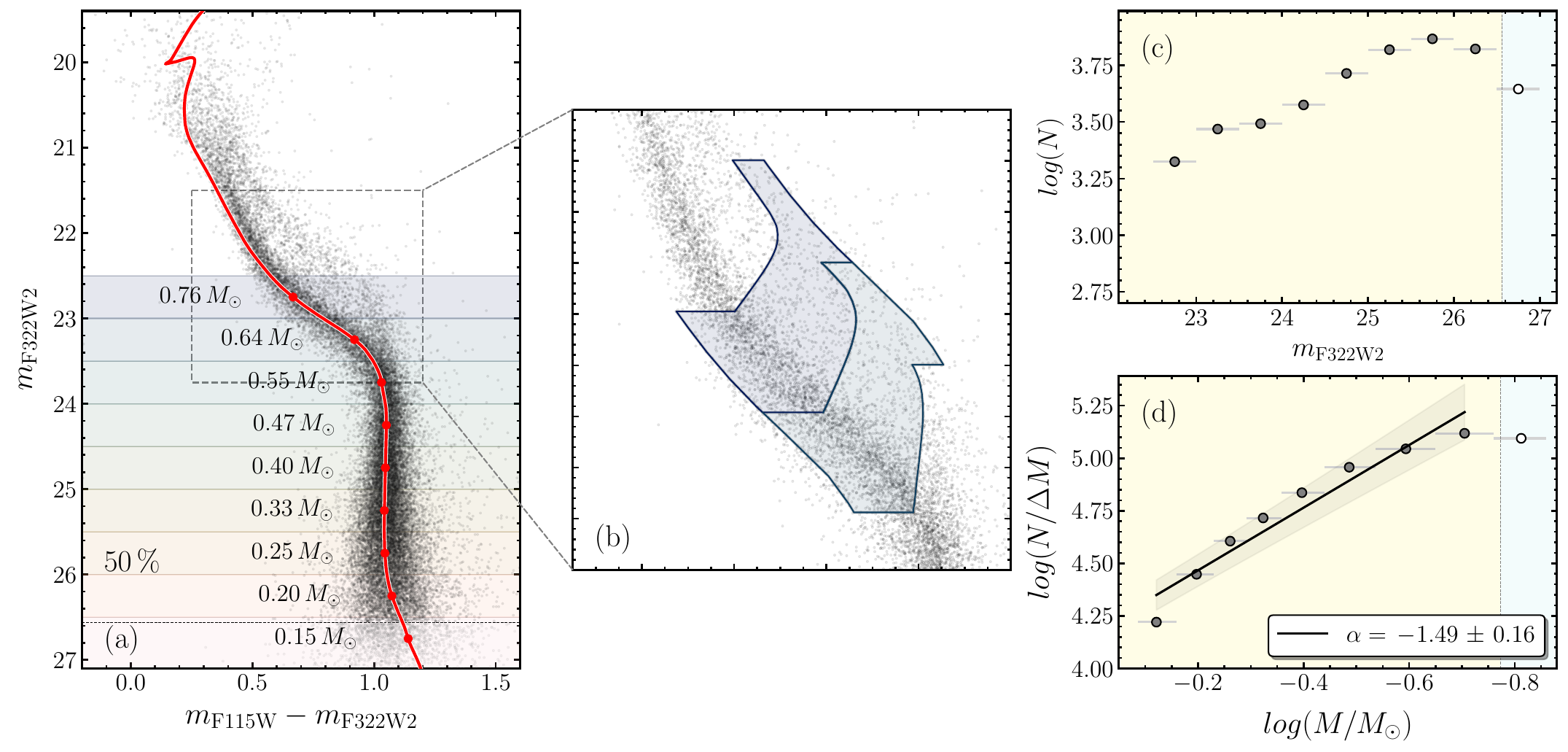}     
    \caption{Determination of the MF of the LMC field analyzed in this work. \textit{Panel a.} $m_{\rm F322W2}$ vs.\,$m_{\rm F115W}-m_{\rm F322W2}$ CMD of LMC stars. The red solid line represents the 2 Gyr BaSTI isochrone adopted to derive the mass–luminosity relation over the magnitude interval $22.5 < m_{\rm F322W2} < 26.6$. The stellar masses associated with the magnitude bins used to derive the MF are marked along the isochrone. {\it Panel b.} Zoomed-in view of the CMD in the magnitude range $21.5 < m_{\rm F322W2} < 23.8$. The two shaded regions mark the bins adopted to derive the MF along the upper MS, constructed to account for the presence of binary systems. {\it Panels c and d.} $m_{\rm F322W2}$ luminosity function (c) and corresponding MF (d) of LMC field stars. In panel d, the black solid line represents a linear fit to the observed MF, restricted to stars in the magnitude range $22.5 <m_{\rm F322W2}<26.6$. The derived MF slope is indicated in the bottom-right corner of the figure. In all panels, the gray dashed line indicates the 50$\%$ completeness limit.}
    \label{fig:LMC_MF_det}
\end{figure*}
%%%%%%%%%%%%%%%%%%%%%%%%%%%%%%%%%
%%%%%%%%%%%%%%%%%%%%%%%%%%%%%%%%%
\begin{figure}
    \centering
    \includegraphics[width=.45\textwidth]{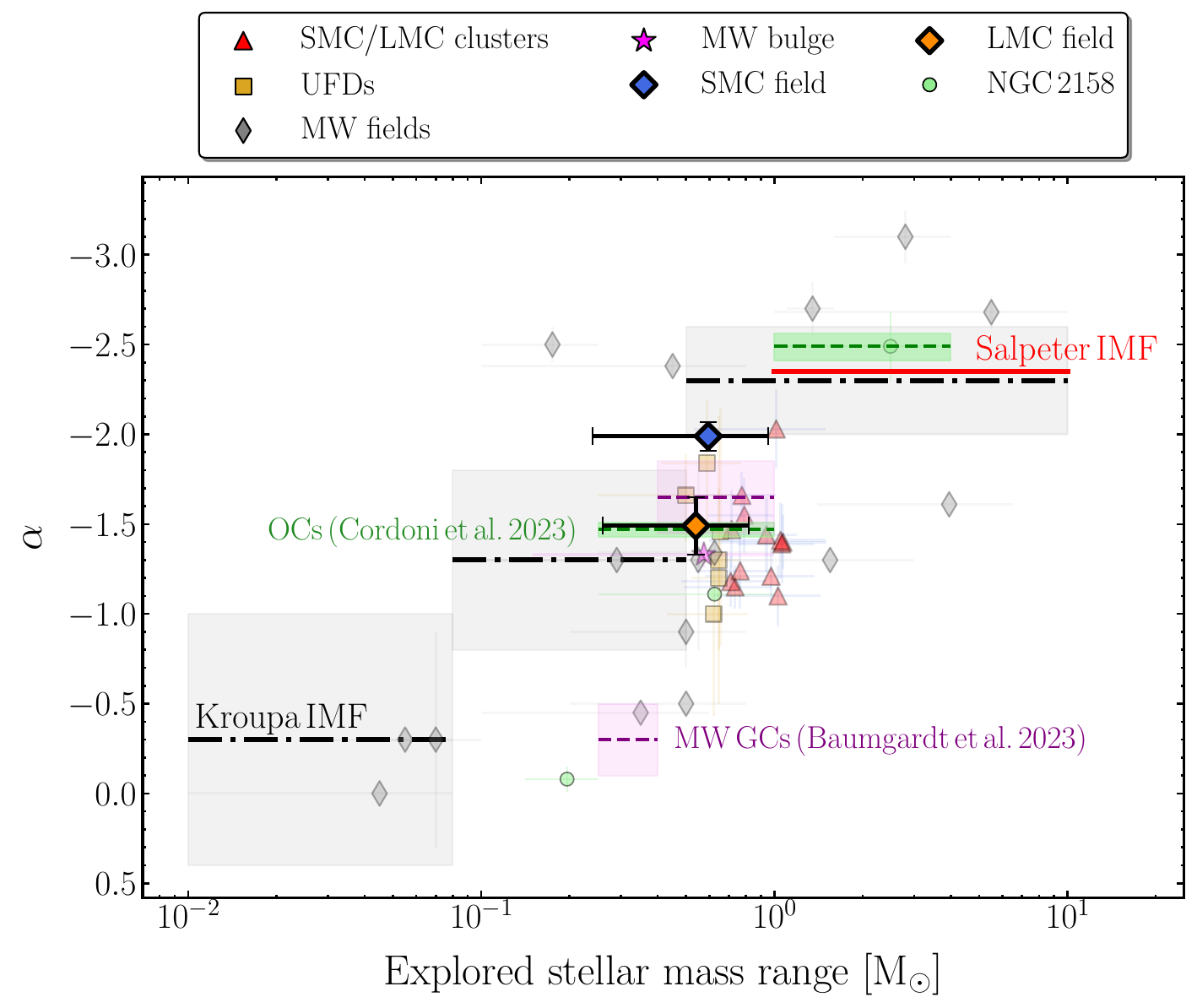}       
    \caption{MF slope, $\alpha$, as a function of the stellar-mass range probed in a variety of Galactic and extragalactic environments. Each symbol shows the best-fitting single power-law slope measured over a given mass interval; horizontal error bars indicate the width of that interval. Red triangles denotes SMC and LMC clusters, gold squares UFDs, gray diamonds Milky Way field stars, the magenta star the Galactic bulge, and the light green points NGC\,2158. The orange diamond highlights the LMC-field measurement from this work, while the blue diamond shows the SMC-field result from \cite{legnardi2025}. The solid red line indicates the canonical \citet{salpeter1955} IMF slope; dashed black lines show the segmented IMF slopes from \citet{kroupa2001}. The green and purple dashed lines mark the slopes reported by \cite{cordoni2023} for Galactic open clusters and by \cite{baumgardt2023} for Milky Way GCs, respectively.}
    \label{fig:alpha_plot}
\end{figure}
%%%%%%%%%%%%%%%%%%%%%%%%%%%%%%%%%

\section{The mass function of field stars in the Large Magellanic Cloud}
\label{sec:MF}
In this section, we derive the low-mass end of the MF of LMC field stars. Owing to the dynamically mixed nature of field populations, the present-day MF is expected to closely trace the IMF over the mass range considered here \citep[see][and references therein]{geha2013}. 

The MF is computed using stars in the magnitude range $22.5<m_{\rm F322W2}<26.6$, corresponding to the region delimited by the horizontal dashed lines in Fig.~\ref{fig:cmd_fit}. This selection avoids (i) the bright portion of the MS where the mass-luminosity relation becomes strongly age-dependent, and (ii) the faint regime where photometric completeness drops below $50\%$.

The MF determination follows two steps. First (Sect.~\ref{subsec:LF}) we measure the luminosity function in $m_{\rm F322W2}$, explicitly accounting for unresolved binaries in the $m_{\rm F322W2}$ versus\,$m_{\rm F115W}-m_{\rm F322W2}$ CMD. Second (Sect.~\ref{subsec:MF}) we convert the luminosity function into a MF using a mass-luminosity relation and fit its slope. 

\subsection{The luminosity function of field stars in the Large Magellanic Cloud}
\label{subsec:LF}
We derived the luminosity function using the binning scheme in Fig.~\ref{fig:LMC_MF_det}, adapted from \citet[][see also \citealt{marchuk2026}]{legnardi2025}. The adopted bins are designed to minimize the mixing of different primary masses induced by unresolved binaries. Along the upper MS, where binaries can be identified more reliably, we defined two CMD regions following Sect.~\ref{sec:bin}. As shown in panel~b of Fig.~\ref{fig:LMC_MF_det}, each region includes single MS stars in a 0.5-mag interval within $22.5 < m_{\rm F322W2} < 23.5$, as well as binaries with $0<q<1$ whose primary falls in the same magnitude interval. For stars with $m_{\rm F322W2}>23.5$, that is, below the MS knee, we adopted a uniform binning scheme with bins of width 0.5 mag.

Photometric errors scatter stars across bin boundaries, producing mutual contamination between adjacent bins. We corrected for this effect using the approach introduced by \citet{milone2012b} and widely applied in subsequent works \citep[][]{dondoglio2022,legnardi2025,marchuk2026}. The completeness-corrected number of stars in the $i$-th bin, $N_i$, is given by
\begin{equation}\label{eq:LF}
    N_i=\sum_k n_kc_{k,i},
\end{equation}
where $n_k$ is the intrinsic number of stars in the $k$-th bin, and $c_{k,i}$ is the contamination matrix element, that is, the fraction of stars from the $k$-th bin that are observed in the $i$-th bin as a consequence of observational uncertainties. We derived $c_{k,i}$ from AS-based synthetic CMDs and obtained the intrinsic counts $n_k$ by solving the linear system defined by Eq.~\ref{eq:LF}.

We also corrected the luminosity function for the contribution of unresolved binaries. Following \citet{legnardi2025}, we adopted the total binary fraction inferred in Sect.~\ref{sec:bin}, $f_{\rm bin}^{\rm TOT}=0.34\pm0.02$, and assumed a flat mass-ratio distribution. For each magnitude bin, unresolved companions were generated by drawing mass ratios uniformly over the range $q_{min}<q<1$, where $q_{min}=\frac{M_{lim}}{M_1}$, $M_1$ is the primary mass, and $M_{lim}=0.17\,M_{\odot}$ is the lower mass limit of our analysis. Accordingly, the fraction of binaries contributing companions above the mass limit was scaled as $f_{\rm eff}=f_{\rm bin}^{\rm TOT}(1-q_{min})$. Secondary masses were converted into F322W2 magnitudes using the mass-luminosity relation of the isochrone shown in Fig.~\ref{fig:LMC_MF_det}a, and the companions were added to the corresponding magnitude bins. This procedure accounts for the contribution of unresolved binaries to the observed luminosity function while avoiding an overestimate of the binary correction in the lowest-mass bins.

The resulting luminosity function is shown in panel~c of Fig.~\ref{fig:LMC_MF_det}. The uncertainties associated with each bin, accounting for both the Poisson statistics of the observed star counts and the uncertainty in the completeness correction, were estimated as $\sigma_{N_i}=\sqrt{\frac{N_i}{C^2_i}+\frac{N_i^2}{C^3_i}\frac{1-C_i}{N_{AS,i}}}$, where $C_i$ is the average completeness factor of the stars in the bin and $N_{AS,i}$ is the number of injected ASs.  

\subsection{The mass function slope in the field of the Large Magellanic Cloud}
\label{subsec:MF}
To convert luminosities into stellar masses, we adopted the BaSTI isochrone shown as a red solid line in Fig.~\ref{fig:LMC_MF_det}a \citep[][]{pietrinferni2021}. The adopted model has age 2\,Gyr, metallicity $Z=0.006$, and $[\alpha/{\rm Fe}]=+0.2$, assuming a distance modulus of $(m-M)_0=18.5$ and a foreground reddening of $E(B-V)=0.01$. 

The resulting MF is displayed in panel~d of Fig.~\ref{fig:LMC_MF_det}. Previous studies of LMC fields \citep[e.g.,][]{gouliermis2005,gouliermis2006}, based on the observations taken with the Wide Field Planetary Camera 2 on board the HST, were limited to $M \gtrsim 0.7\,M_{\odot}$ and therefore could not probe the low-mass regime. Thanks to the depth of the JWST data, we extended the MF down to $M \sim 0.15\,M_{\odot}$, as illustrated in Fig.~\ref{fig:LMC_MF_det}a. For the slope determination, however, we restricted the fit to the portion of the MF where the completeness is higher than 50$\%$, that is, for stars with $m_{\rm F322W2} < 26.56$ and $M > 0.17\,M_{\odot}$.

The MF is commonly described by a power law
\begin{equation}
\frac{dN}{dM} = k \cdot M^{\alpha},
\end{equation}
where $k$ is a normalization constant and $\alpha$ is the slope. In this notation, the canonical \citet{salpeter1955} IMF has a slope of $\alpha=-2.35$, which provides a useful reference for comparison with the MFs derived below. In logarithmic form,
\begin{equation}
\log\left(\frac{dN}{dM}\right) = \log k + \alpha \log M,
\end{equation}
so that $\alpha$ can be obtained from a linear fit in the $\log(dN/dM)$ versus\,$\log M$ plane (Fig.~\ref{fig:LMC_MF_det}d). The best-fitting slope is $\alpha = -1.49 \pm 0.16$ (black solid line).

Adopting a break at $M=0.5\,M_{\odot}$, as in the canonical \cite{kroupa2001} IMF, we obtained slopes of $\alpha=-2.76 \pm 0.15$ for $M>0.5\,M_{\odot}$ and $\alpha=-1.15 \pm 0.11$ for $M<0.5\,M_{\odot}$. The low-mass slope agrees, within the uncertainties, with the \cite{kroupa2001} value, $\alpha=-1.3$, whereas the high-mass slope is steeper than the corresponding Galactic-field estimate of $\alpha=-2.3$ reported by \citet[][]{kroupa2001}.

A single power-law fit over the full mass range yields $\alpha=-1.49 \pm 0.16$. Since the interval considered here ($0.17$-$0.82\,M_{\odot}$) spans the break at $0.5\,M_{\odot}$, a \cite{kroupa2001} IMF approximated by a single power law is expected to have an effective slope of $\alpha\sim-1.5$. The measured global slope is therefore fully consistent with a Kroupa-like IMF, and we adopt the single power-law fit as the fiducial MF in the following analysis. The best-fitting slopes are summarized in Table\,\ref{tab1}.

In Fig.~\ref{fig:alpha_plot} we compare our measurement with MF slopes compiled for a range of Galactic and extragalactic systems. These include Galactic GCs \citep[purple dashed lines;][]{baumgardt2023}, Magellanic Cloud clusters \citep[red triangles;][]{baumgardt2023}, open clusters \citep[green dashed lines;][]{cordoni2023}, UFDs \citep[gold squares;][]{geha2013,gennaro2018a}, Milky Way fields \citep[gray diamonds;][]{reid1999,reid2002,schroder2003,kroupa2002,allen2005,metchev2008,pinfield2008,bochanski2010,sollima2019}, the SMC field \citep[blue diamond;][]{legnardi2025}, the Galactic bulge \citep[magenta starred symbol;][]{zoccali2000}, and the open cluster NGC\,2158 \citep[light green points;][]{marchuk2026}.

Our LMC field measurement (orange diamond) reaches $0.17\,M_{\odot}$ at the 50$\%$ completeness limit, extending to substantially lower masses than most previous extragalactic determinations. This depth enables a robust characterization of the MF in a low-metallicity, dynamically mixed field population.

The best-fitting slope, $\alpha=-1.49\pm0.16$, is shallower than the canonical Salpeter value (red solid line) and slightly shallower than that measured in the SMC field (blue diamond). Moreover, it is consistent, within the uncertainties, with the MF slopes reported for Milky Way GCs over the mass range $0.4$-$1.0\,M_{\odot}$. Conversely, for stars with $M<0.4\,M_{\odot}$, the slope measured in the LMC field is significantly steeper than the value predicted by the parameterization of \cite{baumgardt2023} for Galactic GCs ($\alpha=-0.3\pm0.2$). Over the mass interval explored by our observations, our MF is also consistent with the parameterization derived by \cite{cordoni2023} for Galactic open clusters, as well as with the slopes measured in several Magellanic Cloud clusters. Taken together, these comparisons suggest that environmental conditions play only a minor role in shaping the low-mass MF, while emphasizing that our conclusions are restricted to the mass range sampled by the present observations.

%%%%%%%%%%%%%%%%%%%%%%%%%%%%%%%%%
\begin{table}[ht]
    \centering
        \caption{Inferred parameters of the LMC field. Distance modulus, metallicity, [$\alpha$/Fe], and foreground reddening are derived from isochrone fitting; binary fractions ($f_{\rm bin}^{q>0.6}$ and $f_{\rm bin}^{\rm TOT}$) are obtained from CMD analysis; MF slopes are measured from linear fits in the $\log(N/\Delta M)$–$\log(M)$ plane.}
    \setlength{\tabcolsep}{8pt} 
    \renewcommand{\arraystretch}{1.3} 
    \begin{tabular}{c c }
        \hline
        Parameter & Value \\
        \hline
        $(m-M)_0$ & 18.50 \\
        $Z$ & 0.006 \\
        {[$\alpha$/Fe]} & +0.2 \\
        $E(B-V)$ & 0.01 \\
        $f_{\rm bin}^{q>0.6}$ & $0.15\pm0.01$ \\
        $f_{\rm bin}^{\rm TOT}$ & $0.34\pm0.02$ \\
        $\alpha$ ($0.17$-$0.82\,M_{\odot}$) & $-1.49 \pm 0.16$ \\
        $\alpha$ ($0.17$-$0.50\,M_{\odot}$) & $-1.15 \pm 0.11$ \\
        $\alpha$ ($0.50$-$0.82\,M_{\odot}$) & $-2.76 \pm 0.15$ \\
        \hline
    \end{tabular}
    \label{tab1}
\end{table}
%%%%%%%%%%%%%%%%%%%%%%%%%%%%%%%%%

\section{Summary and discussion}
\label{sec:concl}
In this work, we presented ultra-deep JWST/NIRCam photometry in the F115W and F322W2 bands of the intermediate-age and massive LMC cluster NGC\,1846 and its surrounding field. We separated cluster and field stars using the radial density profile of NGC\,1846. While a forthcoming paper will focus on the properties of the cluster itself, here we exploited the depth and photometric precision of JWST to characterize the stellar population of the outer LMC field. In particular, we investigated the binary-star population and derived the stellar MF down to unprecedentedly low masses. Overall, our results indicate that both binary formation and the low-mass MF are only weakly dependent on the host environment. Our main findings can be summarized as follows:

\begin{itemize}
    \item We measured a fraction of binary systems with mass ratio $q>0.6$ of $f_{\rm bin}^{q>0.6}=0.15\pm0.01$. Assuming a flat mass-ratio distribution over the interval $0<q<1$, this corresponds to a total binary fraction of $f_{\rm bin}^{\rm TOT}=0.34\pm0.02$. This value is consistent with measurements in both the Milky Way field \citep{offner2023} and the SMC field \citep{legnardi2025}. Additionally, thanks to the exceptional angular resolution of JWST/NIRCam, we characterize for the first time the population of wide binaries in the LMC field. Adopting a maximum projected separation of 5 pixels (i.e., 7750\,AU or 0.16\arcsec) between the two components, we measured a negligible fraction of wide binaries with $q>0.6$, $f_{\rm resolved\,bin}^{q>0.6}=0.01\pm0.01$. This value is consistent with previous measurements in other low-density environments, including the open cluster NGC\,2158 and the UFDs Reticulum\,II and Bootes\,I \citep[][]{safarzadeh2022,sharjat2025,marchuk2026}.

    \item We derived the MF of LMC field stars accounting for photometric uncertainties and unresolved binaries. The MF extends down to $0.17\,M_{\odot}$, significantly below the mass range probed in previous studies \citep[e.g.,][]{gouliermis2005,gouliermis2006}. A single power-law fit yields a slope of $\alpha=-1.49\pm0.16$, substantially shallower than the Salpeter value of $\alpha=-2.35$. A broken power-law fit with a fixed break at $M=0.5\,M_{\odot}$ gives $\alpha=-2.76\pm0.15$ for $M>0.5\,M_{\odot}$ and $\alpha=-1.15\pm0.11$ for $M<0.5\,M_{\odot}$, with the low-mass slope consistent with a Kroupa IMF, while the higher-mass slope is steeper than the canonical Galactic value. Overall, the global slope is consistent with the effective slope expected for a Kroupa-like IMF over the mass range sampled by our data. 
\end{itemize}

These findings demonstrate the unique capability of JWST to resolve and characterize faint stellar populations in nearby galaxies, opening a new observational window on the low-mass stellar content of the Magellanic Clouds. Extending similar analyses to other LMC and SMC fields spanning different star-formation histories and environmental conditions will be crucial to establish whether the observed MF variations are universal and to further constrain the role of environment in shaping both the IMF and binary-star populations. 

\section*{Data availability}
Photometry is available at the CDS via anonymous ftp to \url{cdsarc.u-strasbg.fr (130.79.128.5)} or via \url{http://cdsweb.u-strasbg.fr/cgi-bin/qcat?J/A+A/}.

\begin{acknowledgements}
We thank the anonymous referee for various suggestions that improved the quality of the manuscript. This work is based on observations made with the NASA/ESA/CSA James Webb Space Telescope. The data were obtained from the Mikulski Archive for Space Telescopes at the Space Telescope Science Institute, which is operated by the Association of Universities for Research in Astronomy, Inc., under NASA contract NAS 5-03127 for JWST. These observations are associated with program GO-9012. A.\,B. acknowledges support from STScI grant GO-9012. E.\,P.\,L. acknowledges support by Special Project for High-End Foreign Experts "Xingdian" Funding from Yunnan Province, and National Key R\&D Program of China Grant (No. 2024YFA1611601). S.\,J. acknowledges support from the National Research Foundation of Korea (NRF) funded by the Ministry of Education (RS-2025-25419519 and RS-2022-NR070872) and by the NRF grant funded by the Korea government (MSIT) (RS-2022-NR070525).
\end{acknowledgements}

\bibliographystyle{aa}
\bibliography{example}
\end{document}